\documentclass{iopart}
\def\l{\label}
\def\ct{\cite}
\def\r{\ref}

\def\ep{\varepsilon}
\def\rd{\displaystyle{\cdot}}

\def\d{\mbox{d}}
\def\D{\mbox{D}}

\def\la{\langle}
\def\ra{\rangle}
\def\tl{\tilde}
\def\ts{\textstyle}
\def\c{\mbox{curl}\,}
\def\hsp5{\hspace{5mm}}
\newcommand{\sfrac}[2]{{\textstyle{#1\over#2}}}
\def\case#1/#2{\textstyle\frac{#1}{#2}}
\def\apj{{\em Astrophys. J.\/} }

\def\cqg{{\em Class. Quantum Grav.\/} }

\def\mn{{\em Mon. Not. R. Astron. Soc.\/} }
\def\prd{{\em Phys. Rev.\/} D }

\begin{document}
\jl{6}
\title{General relativistic analysis of peculiar velocities}
\author{
George F. R. Ellis$^{1,2}$\footnote[1]{E-mail: {\tt
ellis@maths.uct.ac.za}}, Henk van Elst$^2$\footnote[2]{E-mail:
{\tt H.van.Elst@qmul.ac.uk}} and Roy
Maartens$^3$\footnote[3]{E-mail: {\tt Roy.Maartens@port.ac.uk}}}

\address{~}
\address{$^1$Cosmology Group, Department of Mathematics and
Applied Mathematics, University of Cape Town, Rondebosch~7701,
South Africa}
\address{$^2$Astronomy Unit, Queen Mary, University of London,
Mile End Road, London~E1~4NS, United Kingdom}
\address{$^3$Relativity and Cosmology Group, School of Computer
Science and Mathematics, Portsmouth University,
Portsmouth~PO1~2EG, Britain}

\date{August 30, 2001}
\begin{abstract}
We give a careful general relativistic and (1+3)-covariant
analysis of cosmological peculiar velocities induced by matter
density perturbations in the presence of a cosmological constant.
In our quasi-Newtonian approach, constraint equations arise to
maintain zero shear of the non-comoving fundamental worldlines
which define a Newtonian-like frame, and these lead to the
(1+3)-covariant dynamical equations, including a generalized
Poisson-type equation. We investigate the relation between
peculiar velocity and peculiar acceleration, finding the
conditions under which they are aligned. In this case we find
(1+3)-covariant relativistic generalizations of well-known
Newtonian results.

\end{abstract}
\pacs{04.20.-q, 98.80.Hw, 98.65.Dx, 95.30.Sf~\hfill
{gr-qc/0105083}}
\submitted

\section{INTRODUCTION}
Density and velocity perturbations of cold matter models of the
observable part of the expanding Universe are central to the
analysis of structure formation in cosmology (see e.g. the recent
review by Dekel~\cite{d}). On scales well below the Hubble radius,
the Newtonian theory of gravitation is a good approximation, and
Peebles' approach~\cite{pee76} is widely used (see e.g.
\cite{llpr,berdek89,pad93,pea99}). However, observations and
simulations are probing scales which are a significant fraction of
the Hubble radius, thus requiring a relativistic treatment of the
gravitational dynamics in a cosmological context.  Furthermore,
observational evidence for a positive cosmological constant (see
e.g. Perlmutter {\em et al\/}~\cite{peretal98}) implies that this
constant does affect the evolution of matter density and velocity
perturbations on sufficiently large scales.

Relativistic gauge-invariant perturbations of
Friedmann--Lema\^{\i}tre models are usually studied via a
metric-based approach~\cite{bar80}. Bardeen's quasi-Newtonian (or
longitudinal) gauge is well adapted to study the evolution of
matter perturbations~\cite{bar80,muketal92,ber96}. Such an
approach has recently been used to find relativistic corrections
to the Newtonian relation between the peculiar velocities of
matter structures and the matter density
contrast~\cite{manrah99,r}. Here we follow an alternative
relativistic approach, based on the (1+3)-covariant analogue of
the quasi-Newtonian gauge, which was developed in~\cite{hveell98}.
The quasi-Newtonian gauge is based on time slices with zero-shear
normals. The (1+3)-covariant analogue uses a 4-velocity field
$u^{a}$ (or a threading of spacetime) which is irrotational and
shearfree, and so effectively mimics the gauge fixing conditions
of the quasi-Newtonian gauge. Analysis of the evolution and
constraint equations shows that consistency conditions arise from
enforcing vanishing shear and vorticity~\cite{hveell98}. These
consistency conditions lead to the perturbation equations.

Vanishing vorticity reflects the absence of vector modes, while
vanishing vorticity and shear together imply vanishing
gravito-magnetic Weyl curvature, so that tensor modes are also
absent. The remaining, scalar, modes describe peculiar velocity
and matter density perturbations. Peculiar velocity in the matter
distribution is directly generated by inhomogeneity in the
expansion rate of $u^{a}$, which also accounts for time-dependence
of the peculiar gravitational potential. The peculiar velocity and
acceleration satisfy a coupled pair of evolution equations. The
peculiar gravitational potential obeys a generalized Poisson-type
equation, which on small scales reduces to the spatial gradient of
the usual Poisson equation.

The peculiar velocity is {\em not\/} in general aligned with the
peculiar acceleration, but its orthogonal component will remain
zero if it vanishes initially. In general, the magnitude of the
orthogonal component decays with expansion, but in order to keep
this component strictly zero, we arrive at an equation determining
the ratio of peculiar velocity to peculiar acceleration. This
leads to (1+3)-covariant relativistic generalizations of
well-known Newtonian relations, including a cosmological constant.

\section{(1+3)-COVARIANT QUASI-NEWTONIAN ANALYSIS}
Peculiar velocities must be defined relative to a preferred frame
with 4-velocity field $u^{a}$. This preferred frame is necessarily
{\em non-comoving\/} relative to the matter, since, by definition,
peculiar velocities vanish in the comoving (Lagrangian) frame.
(Much of the standard literature on structure formation uses the
term ``comoving'' in a rather misleading fashion, i.e. as comoving
with the {\em fictitious\/} background 4-velocity rather than with
the actual matter 4-velocity.) A (1+3)-covariant (and local)
approach seeks to define the preferred frame in a physical way.
(For a recent review on (1+3)-covariant methods see
e.g.~\cite{ellhve99}; for a detailed discussion of the relation
between the (1+3)-covariant and metric-based approaches
see~\cite{bruetal92}.) Motivated by the Newtonian analysis of
Peebles~\cite{pee76}, we choose $u^a$ to be irrotational
($\omega^{a} = 0$) and shearfree ($\sigma_{ab}=0$), defining a
quasi-Newtonian frame~\cite{hveell98} corresponding to Bardeen's
quasi-Newtonian gauge~\cite{bar80}. Furthermore, we require that
$u^a$ is non-relativistic relative to observers comoving with the
matter, whose 4-velocity field $\tl{u}^{a}$ is given by
\begin{equation}
\label{tuu}
\tl{u}^{a} = u^{a} + v^{a} \ , \quad v_{a}u^{a} = 0 \ , \quad
v_{a}v^{a} \ll 1 \ .
\end{equation}
Here $v^{a}$ is the non-relativistic peculiar velocity, which
vanishes in the background (and hence is gauge-invariant).

Thus we have {\em two\/} physical frames, defined by the
(pressure-free) matter and by the quasi-Newtonian observers. In
the comoving (Lagrangian) frame $\tilde{u}^a$, the pressure,
energy current density, anisotropic pressure and 4-acceleration
all vanish:
\begin{equation}
\tilde{p} = \tilde{q}{}^{a} = \tilde{\pi}_{ab} = 0 \ , \quad
\tilde{A}{}^{a} = 0 \ .
\end{equation}
In the quasi-Newtonian (Eulerian) frame $u^a$,
\begin{equation}
\label{2}
\omega^{a} = \sigma_{ab} = 0 \ .
\end{equation}
The Galilean velocity boost in (\ref{tuu}) preserves the pressure
and energy density to linear order in $v^{a}$, i.e., $p =
\tilde{p}$, $\mu = \tilde{\mu}$, but introduces a non-zero energy
current density (momentum density) in the Eulerian frame given by
\begin{equation}
\label{1}
q^{a} = \rho\,v^{a}  \ ,
\end{equation}
where $\rho$ is the background matter density. Also to linear order
in $v^{a}$ we have $\pi_{ab} = 0$. (These linear results are
obtained from the exact boost transformation laws (B7) -- (B10)
given in~\ct{maa98}.)  In addition, it transforms the kinematic
quantities to linear order in $v^{a}$ as follows (cf.~the exact
boost transformation laws (B3) -- (B6) in~\ct{maa98}):
\begin{eqnarray}
\label{7}
\tl{\Theta} & = & \Theta + \D_{a}v^{a} \ , \\
\label{8}
\tl{A}^{a} & = & A^{a} + \dot{v}{}^{a} + H\,v^{a} = 0 \ , \\
\label{9}
\tl{\omega}{}^{a} & = & \omega^{a} - {\ts{1\over2}}\,(\c
v)^{a} = -{\ts{1\over2}}\,(\c v)^{a} \ , \\
\label{10}
\tl{\sigma}_{ab} & = & \sigma_{ab} + \D_{\la a}v_{b\ra}
= \D_{\la a}v_{b\ra} \ .
\end{eqnarray}
Notation and some fundamental equations are given in the
appendix. The gravito-electric/-magnetic Weyl curvature fields are
frame-invariant to linear order in $v^{a}$ (cf.~the exact boost
transformation laws (B11) and (B12) in~\ct{maa98}):
\begin{equation}
\label{14}
\tl{E}_{ab} = E_{ab} \ , \quad \tl{H}_{ab} = H_{ab} \ .
\end{equation}
The gravito-magnetic constraint equation (\ref{a8}), together with
(\ref{2}) and (\ref{14}), shows that
\begin{equation}
\label{2'}
H_{ab} = 0 = \tilde{H}_{ab} \ ,
\end{equation}
which reflects the absence of tensor modes (gravitational
radiation). The div-$H_{ab}$ constraint equation (\ref{a10}),
together with (\ref{2}) and (\ref{2'}), shows that $q^{a}$ is
irrotational, and thus so is $v^{a}$:
\begin{equation}
\label{2''}
(\c q)^{a} = 0 \quad \Rightarrow \quad (\c v)^{a} = 0
\ .
\end{equation}
Thus
\begin{equation}
\tl{\omega}{}^{a} = 0 \ ,
\end{equation}
by (\ref{9}), reflecting the absence of vector (rotational) modes.
Only scalar modes are admitted, and for these
\begin{eqnarray}
\label{scalar1}
&& W_a = \D_a W \ , \quad (\c W)^{a} = 0 \ , \\
\label{scalar2}
&& S_{ab} = \D_{\la a}\D_{b\ra}S \ , \quad (\c
S)_{ab} = 0 \ ,
\end{eqnarray}
where $W^{a}$ is a vector orthogonal to $u^a$, $S_{ab}$ is a
tracefree tensor orthogonal to $u^a$, and $W$ and $S$ are scalars.
In particular (to linear order in the second case),
\begin{equation}
\label{pot} A_{a} = \D_{a}\Phi \ , \quad v_{a} =
\D_{a}\left(\frac{2\Theta}{3\rho}\right) \ ,
\end{equation}
where $\Phi$ is the peculiar gravitational potential, whose
existence follows from (\ref{2}) and the vorticity evolution
equation (\ref{a2}). The velocity potential follows from (\ref{1})
and the div-$\sigma_{ab}$ constraint equation (\ref{a6}). (Note
that these potentials have a trivial gauge freedom to add an
arbitrary homogeneous function.) The velocity potential is
exploited in the POTENT method of Bertschinger and
Dekel~\cite{berdek89} to ``reconstruct'' $v^{a}$ from
observational data. The shear propagation equation (\ref{a3}) then
shows that in the present context the gravito-electric Weyl
curvature is a purely tidal field determined by the peculiar
gravitational potential,
\begin{equation}
\label{tidal}
E_{ab} = \D_{\la a} \D_{b\ra}\Phi \ ,
\end{equation}
in direct correspondence to the Newtonian situation \cite{ell71}
(see also~\cite{bruetal92}).

\section{PERTURBATION EVOLUTION}
In the (1+3)-covariant approach based on a 4-velocity field
$u^{a}$, because of the expanding Universe context, the
relativistic equations do not directly provide an analogue of the
Newtonian Poisson equation for the peculiar gravitational
potential. Rather, the equations governing the peculiar
gravitational potential are integrability conditions, which arise
directly from the fact that, with $\sigma_{ab} = 0$, the shear
propagation equation (\ref{a3}) is turned into a {\em new\/}
constraint equation, i.e., ${\cal E}_{ab}
\equiv E_{ab}-\D_{\la a} A_{b\ra} = 0$. The time evolution
$\dot{\cal E}_{ab}$ and the spatial divergence $\D_{b}{\cal
E}^{ab}$ of this constraint equation give us the dynamical
equations for the peculiar gravitational potential, which are (in
units where $c = 1 =8\pi G/c^{2}$)~\cite{hveell98,maa98}
\begin{eqnarray}
\label{q3}
&& \dot{\Phi} = - {\ts{1\over3}}\,\Theta \ , \\
\label{new}
&& \D_a\left(\D^2\Phi\right) = {\ts{1\over2}}\,\D_a\mu
- 3\,{{\cal K}\over a^2}\,\D_a\Phi - H\,\D_a\Theta \ ;
\end{eqnarray}
Here $\D^{2}$ is the covariant Laplace operator for a space of
constant curvature ${\cal K}/a^2$, with ${\cal K} = 0, \,\pm 1$,
and $a$ is the background scale factor. Note in particular that
(\ref{q3}) is {\em not\/} gauge-invariant, but we can use it to
rewrite the gauge-invariant equation (\ref{new}) as
\begin{equation}
\label{new'}
\D_a\left[\ \D^2\Phi-3H\dot{\Phi}+3{{\cal K}\over
a^2}\Phi\ \right] = \D_a\left[\ {\ts{1\over2}}\,\mu\ \right] \ ,
\end{equation}
which can be immediately integrated to yield
\begin{equation}
\D^2\Phi-3H\dot{\Phi}+3{{\cal K}\over
a^2}\Phi = {\ts{1\over2}}\,\delta\mu \ ,
\end{equation}
where we have used the gauge freedom in $\Phi$, i.e.,
$\Phi\to\Phi+\beta(t)$, to remove the background part $\rho$ of the
matter density $\mu$ on the right. This is the (1+3)-covariant
relativistic generalization of the Newtonian Poisson equation
$\vec{\nabla}^{2}\Phi = {1\over2}\,\delta\mu$. The spatial
gradients in (\ref{new'}) ensure gauge-invariance, since each term
vanishes in the background (note that $\delta\mu$ is
gauge-dependent). There are two relativistic corrections: one to
incorporate background spatial curvature ${\cal K}$, and one to
incorporate time evolution of the peculiar gravitational
potential. The latter is directly induced by the peculiar velocity
of the matter. [\,An alternative relativistic Poisson equation is
given in the metric-based approach by Bardeen~\cite{bar80},
Eq.~(4.3).\,]

In the linear regime of structure formation, the time scale of
variations in $\Phi$ is of the order of the Hubble time $H^{-1}$,
and on a length scale $\lambda$, we have
$\D^2\Phi\sim\Phi/\lambda^2$. Thus, if $\lambda\ll H^{-1}$, i.e.,
on scales well within the Hubble radius, we have
\begin{equation}
|H\dot\Phi|\sim H^2|\Phi| \ll |\D^2\Phi|\sim {|\Phi|\over
\lambda^2} \ .
\end{equation}
On these scales, where background spatial curvature effects are
also negligible, we recover the spatial gradient of the Poisson
equation.

\subsection{PECULIAR VELOCITY AND PECULIAR ACCELERATION}
Taking the spatial gradient of (\ref{q3}), and using the
div-$\sigma_{ab}$ constraint equation (\ref{a6}), we get
\begin{equation}
\label{q4'}
\dot{A}{}^{a} + 2\,H\,A^{a} = - {\ts{1\over2}}\,\rho\,v^{a} \ ,
\end{equation}
the evolution equation for $A^{a}$. This is coupled to the Euler
momentum equation (\ref{c4}) yielding the evolution equation for
$v^{a}$ as
\begin{equation}
\label{q17}
\dot{v}{}^{a} + H\,v^{a} = - A^{a} \ .
\end{equation}
The source term in (\ref{q17}), given by the peculiar acceleration
$A_{a}= \D_{a}\Phi$, shows how peculiar velocity tends to be
generated by the spatial gradient of the peculiar gravitational
potential, which is in turn generated by matter over- and
under-densities through a generalized Poisson-type equation derived
below.

It is reasonable to assume that the peculiar velocity is aligned
with the peculiar acceleration because it is generated by that
acceleration~\ct{pee76}, and indeed this assumption is part of the
Zel'dovich approximation in the weakly non-linear
regime~\ct{zel70}. However, in general there will be an orthogonal
component, i.e.,
\begin{equation}
\label{fw}
v^{a} = -FA^{a} + w^{a} \ ,
\quad A_{a}w^{a} = 0 \ ,
\end{equation}
with $F$ a `growth factor'. Substituting into (\ref{q17}), and
using (\ref{q4'}), we get
\begin{equation}
\label{fw'}
\dot{w}{}^{a} + (H+{\ts{1\over2}}\rho F)\,w^{a}
= \left[\dot{F} - (H-\sfrac{1}{2}\rho F)\,F - 1\right]A^{a}  \ .
\end{equation}
In order to maintain $A_{a}w^{a} = 0$ relative to the
quasi-Newtonian frame, we find that the `growth factor' $F$ must
evolve according to
\begin{equation}
\l{fdot1}
\dot{F} = (H-\sfrac{1}{2}\rho F)\,F + 1 +
\left(\frac{\rho}{2A^{2}}\right)w^{2} \ ,
\end{equation}
where $w^{2} = w_{a}w^{a}$ and $A^{2} = A_{a}A^{a}$. This equation
shows, as expected for scalar perturbations, that it is impossible
to have a purely orthogonal peculiar velocity ($F = 0$). If we
contract (\r{fw'}) with $w_{a}$, we find the evolution equation for
$w$:
\begin{equation}
\label{w}
\dot{w} = -\,2\,(H+{\ts{1\over2}}\rho F)\,w \ .
\end{equation}
It follows that if $w^{a} = 0$ at some initial time $t_0$, then
$w^{a}$ remains zero for all $t>t_0$: irrotational gravitational
instability does not generate peculiar velocities orthogonal to the
peculiar acceleration (i.e., to the gradient of the peculiar
gravitational potential). If there {\em is\/} an initial orthogonal
component $w^{a}$ due to random motions, then expansion and matter
aggregation serve to decrease it. Furthermore, the rate of change
of direction of any initial orthogonal component is always along
the spatial gradient of $\Phi$: writing $w^{a} = w\,e^{a}$, where
$e_{a}e^{a} = 1$, we find from (\ref{fw'}) and (\ref{w}) that
\begin{equation}
w\,\dot{e}{}^{a} = \left[\,\dot{F} - (H-\sfrac{1}{2}\rho F)\,F
- 1\,\right]A^{a} \ .
\end{equation}
We are thus justified in assuming that $w^{a} = 0$. In this case,
\begin{equation}
\label{ffw} v^{a} = -FA^{a} \ ,
\end{equation}
where $F$ directly relates $v^{a}$ and $A^{a}$.  Then (\ref{fdot1})
shows that $F$ is a solution of
\begin{equation}
\l{fdot2}
\dot{F} = (H-\sfrac{1}{2}\rho F)\,F + 1 \ ,
\end{equation}
while $v^{a}$ follows from (\ref{q17}), using
(\ref{ffw}), as
\begin{equation}
\label{v}
v^{a} = {\cal V}^{a}\exp\int\left({1\over
F}-H\right)\d t \ , \quad \dot{\cal V}{}^{a} = 0 \ .
\end{equation}
Here ${\cal V}^{a}$ is the initial peculiar velocity field, and $t$
is proper time along $u^a$ in the background. Integration of the
background equation (\ref{fdot2}) to find $F$ will thus determine
the change with time of the peculiar velocity field $v^{a}$ and of
the spatial gradient $\D_{a}\Phi = - F^{-1}v_{a}$. Then (\ref{a9})
gives the corresponding $\D_{a}\mu$.

It is important when solving the Einstein field equations to check
that all the field equations and their associated consistency
conditions are satisfied. This has been verified for the above
approach in the linearised case in~\cite{hveell98}.

\subsection{DENSITY, VELOCITY AND POTENTIAL PERTURBATIONS}
The matter density perturbations are (1+3)-covariantly described
by~\cite{ellbru89,bruetal92}
\begin{equation}
\Delta = a\D_{a}\Delta^{a} \ , \quad \Delta_a =
{a\over\rho}\,\D_a\mu \ ,
\end{equation}
and are given in the present case by~\cite{maa98}
\begin{equation}
\Delta = \tilde{\Delta} + 3a^2H\D_{a}v^{a} \ ,
\end{equation}
where $\tilde{\Delta}$ are the matter density perturbations in the
comoving frame. The latter satisfy the evolution equation
\begin{equation}
\label{dens}
(\tilde{\Delta})^{\rd\rd} + 2H(\tilde{\Delta})^{\rd}
- {\ts{1\over2}}\rho\tilde{\Delta} = 0 \ ,
\end{equation}
which can be solved in a given background. Then the matter density
perturbations follow, using (\ref{v}), as
\begin{equation}
\label{dens2}
\Delta = \tilde{\Delta} + 3aH(\D_{a}{\cal V}^{a})
\exp\left[\int{\d t\over F}\right] \ .
\end{equation}
The usual theory of structure formation in the expanding Universe
context follows (see~\cite{ellbru89,ellhve99}, and references
therein).

By analogy with the Newtonian relation~\cite{pee76,llpr,pea99}
\begin{equation}
\vec{v} = {2Hf\over\rho}\,\vec{g} \ ,
\end{equation}
we define the position-independent dimensionless growth rate $f =
\rho F/2H$ so that
\begin{equation}
\label{linrel}
v^{a} = -\,{2Hf\over\rho}\,A^{a} \ .
\end{equation}
Note that, by (\ref{pot}), we can then write
\begin{equation}
\label{pot'} \D_{a}\left[\Phi+{1\over f}\, {\Theta\over3H}\right]
= 0 \ .
\end{equation}
We can rewrite (\ref{fdot2}) to become part of the dynamical
system
\begin{eqnarray}
\label{f'}
f^{\prime} & = &
-\left(1-\sfrac{1}{2}\,\Omega+\Omega_{\Lambda} +f\right)f
+ \sfrac{3}{2}\,\Omega \ , \\
\Omega^{\prime} & = &
-\left(1-\Omega+2\,\Omega_{\Lambda}\right)\Omega \ , \\
\label{omegalpr}
\Omega_{\Lambda}^{\prime} & = &
2\left(1+\sfrac{1}{2}\,\Omega -\Omega_{\Lambda}\right)
\Omega_{\Lambda} \ .
\end{eqnarray}
This evolves $f$ together with the background variables $\Omega$
and $\Omega_{\Lambda}$, the latter representing the dynamical
effects of the cosmological constant. A prime here denotes a
derivative with respect to the logarithmic dimensionless time
variable $N = \ln(a/a_0) = -\,\ln(1+z)$, where $z$ is the
redshift. Note that by the Friedmann equation (\ref{frw1'}),
$\Omega_{\cal K} = 1 - \Omega - \Omega_{\Lambda}$. Equation
(\ref{v}) now becomes
\begin{equation}
\label{v'}
v^{a} = {\cal V}^{a}\exp\int {(2f-3\Omega)\over2(1+z)f}\,\d z \ ,
\end{equation}
while (\ref{dens2}) becomes
\begin{equation}
\label{dens3}
\Delta = \tilde{\Delta} + 3aH(\D_{a}{\cal V}^{a})
\exp\left[-\int{3\Omega\over 2(1+z)f}\,\d z\right] \ .
\end{equation}
We can use numerical solutions for the growth rate $f$ obtained
from (\ref{f'}) -- (\ref{omegalpr}) in order to evaluate the
integrals in the last set of equations, which are the
(1+3)-covariant relativistic generalizations of well-known
relations derived using Newtonian theory. Within the latter
context a good approximation to $f$ at $z = 0$, with a cosmological
constant, is given by Lahav {\em et al\/}~\cite{llpr} as
\begin{equation}
\label{fa}
f_{0} \approx \Omega_0^{0.6}
+ {{1\over70}}\left(1+{\Omega_0\over2}\right)\Omega_{\Lambda 0} \ .
\end{equation}
(See Hamilton~\cite{ham00} for refinements of this approximation.)

The time-derivative term in the generalized Poisson equation
(\ref{new'}) can be rewritten, using the spatial gradient of
(\ref{q3}) and (\ref{a6}), to give
\begin{equation}
\label{new''}
\D_{a}\left[\,\D^2\Phi-3H^2\left(f-1+\Omega+\Omega_\Lambda\right)
\Phi\,\right]= \D_a\left[\,{\ts{1\over2}}\,\mu\,\right] \ .
\end{equation}
As before, this equation again integrates to give a Poisson-type equation
\begin{equation}
\label{helm} \D^{2}\Phi - 3H^{2} \left(f-\Omega_{\cal
K}\right)\Phi = {\ts{1\over2}}\,\delta\mu \ ,
\end{equation}
on using the gauge freedom in $\Phi$ [\,$\Phi\to\Phi+\beta(t)$\,]
to remove the background part $\rho$ of $\mu$ on the right. The
contrast with the Newtonian equation is obvious.

\subsection{SOLUTIONS}
The solution of (\ref{helm}) may be found using the Green's
function in a constant curvature space for the inhomogeneous
Helmholtz equation. In the case of a spatially flat background,
i.e., when $\Omega _{\cal K} = 0 = 1 - \Omega - \Omega_{\Lambda}$,
this Green's function is given in \cite{aw} and leads to
\begin{equation}
\label{helm'} \Phi(t,\vec{r}\,) = -\,{1\over8\pi}\int
{\delta\mu(t,\vec{r}\,^{\prime})
\over|\vec{r}-\vec{r}\,^{\prime}|}\,
\exp\left[\,-\,H(t)\,\sqrt{3f(t)}\,|\vec{r}-\vec{r}\,^{\prime}|
\,\right]\d^3\vec{r}\,^{\prime} \ .
\end{equation}
With (\ref{linrel}) and (\ref{pot}) this determines the peculiar
velocity field $v^{a}$. The exponential term represents the
relativistic correction to the Newtonian formula. Note that the
effect of the cosmological constant occurs {\em implicitly\/} via
the time-dependence of $H$ and $f$. The way the Newtonian formulae
work out in practice is shown in the graphs given by Lahav {\em et
al\/}~\cite{llpr}.

On scales $\lambda=|\vec{r}-\vec{r}\,^{\prime}|$ well within the
Hubble radius, i.e., $\lambda\ll H^{-1}$, (\ref{helm'}) shows that
\begin{equation}
\Phi\approx\Phi_{\rm Newt}\left(1-\sqrt{3f}H\lambda\right) \ .
\end{equation}

For the Einstein--de Sitter case, i.e., when $\Omega _{\cal K} = 0
= \Omega_{\Lambda}$, (\ref{f'}) has one particular constant
solution given by $f = 1$ which corresponds to a growing mode:
\begin{equation}
v^{a} = {\cal V}^{a} a^{1/2} \ , \quad \Delta = \tilde{\Delta} +
3a^{5/2}H\,(\D_{a}{\cal V}^{a}) = Ca \ ,
\end{equation}
where $\dot{C} = 0$ and we used $\tilde{\Delta} \propto a$.

\section{CONCLUSIONS}
In this paper we discussed a transparent geometrical framework,
based on (1+3)-covariant general relativistic methods, to describe
peculiar velocities arising from inhomogeneity over cosmologically
significant distance scales, that also takes into account a
cosmological constant. Peculiar velocities are defined relative to
a preferred frame. In our case, this is defined by the geometrical
conditions $\omega^{a} = \sigma_{ab} = 0$ imposed on a non-comoving
4-velocity field $u^a$. These relations are always true when
Bardeen's quasi-Newtonian gauge is used to describe the
gravitational dynamics of scalar perturbations. In the
(1+3)-covariant approach the perturbation equations follow from a
consistency analysis of the constraint equations. A relativistic
analysis provides a theoretical basis for determining the peculiar
velocity field on cosmologically significant distance
scales. Conceivable applications of the present framework lie in
analyses such as the POTENT project (see~\ct{berdek89}) to
determine large-scale streaming motions of galaxies within clusters
and superclusters.

A relativistic analysis can also be applied to determine the effect
of peculiar velocities of matter structures on the temperature
anisotropy in the cosmic microwave background radiation, requiring
a harmonic analysis of the above variables and equations. This was
not carried out in the present work, but has been treated in a
(1+3)-covariant approach by Gebbie {\em et
al\/}~\cite{geb99,geb2000,gebetal2000}.

\ack
We acknowledge helpful discussions with Tim Gebbie and Marco
Bruni. HvE was in part supported by the Deutsche
Forschungsgemeinschaft (DFG), Bonn, Germany.


\appendix
\section{Basic equations}
The (1+3)-covariant spatial curls are
\[
(\c V)^{a} = \ep^{abc}\D_{b}V_{c} \ , \quad (\c S)_{ab} =
\ep_{cd(a}\D^cS_{b)}{}^d \ ,
\]
where $\D_{a}$ is the spatially projected covariant derivative and
$\ep_{abc}$ is the spatially projected alternating tensor. The
kinematic quantities are $\Theta = \D_{a}u^{a}$ (rate of
expansion), $A^{a} = \dot{u}{}^{a}$ (4-acceleration), $\omega^{a}
= - {\ts{1\over2}}(\c u)^{a}$ (vorticity), and $\sigma_{ab} =
\D_{\la a}u_{b\ra }$ (rate of shear). The dot denotes the directional
covariant derivative $u^a\nabla_a$ and the angle brackets denote
the projected symmetric tracefree part: $S_{\la ab\ra}\equiv
[h_{(a}{}^ch_{b)}{}^d-{\ts{1\over3}}h^{cd} h_{ab}]S_{cd}$. The
gravito-electric/-magnetic Weyl curvature fields are
\begin{equation*}
E_{ab}=C_{acbd}u^cu^d=E_{\la ab\ra } \ , \quad
H_{ab}={\ts{1\over2}}\ep_{acd}C^{cd}{}{}_{be}u^e=H_{\la ab\ra} \ ,
\end{equation*}
where $C_{abcd}$ is the Weyl curvature tensor. The linearized
(1+3)-covariant evolution equations for dust ($p = \pi_{ab} = 0$)
are (in any frame, using units $c = 1 = 8\pi G/c^{2}$, and
including scalar, vector and tensor modes)~\cite{hveell98,maa98}
\begin{eqnarray}
\label{c3}
\dot{\mu} + \D_{a}q^{a}
& = & - \Theta\,\mu \ , \\
\label{a1}
\dot{\Theta} - \D_{a}A^{a}
& = & - {\ts{1\over3}}\,\Theta^{2}
- {\ts{1\over2}}\,\mu + \Lambda \ , \\
\label{c4}
\dot{q}{}^{a}
& = & - 4\,H\,q^{a} - \rho\,A^{a} \ , \\
\label{a2}
\dot{\omega}{}^{a} + {\ts{1\over2}}(\c A)^{a}
& = & - 2\,H\,\omega^{a} \ , \\
\label{a3}
\dot{\sigma}_{ab} - \D_{\la a}A_{b\ra}
& = & - 2\,H\,\sigma_{ab} - E_{ab} \ , \\
\label{a4}
\dot{E}_{ab} - (\c H)_{ab} + {\ts{1\over2}}\D_{\la a}q_{b\ra}
& = & - 3\,H\,E_{ab} - {\ts{1\over2}}\,\rho\,\sigma_{ab} \ , \\
\label{a5}
\dot{H}_{ab} + (\c E)_{ab}
& = & - 3\,H\,H_{ab} \ ,
\end{eqnarray}
and the linearized (1+3)-covariant constraint equations are
\begin{eqnarray}
\label{a7}
0 & = & \D_{a}\omega^{a} \ , \\
\label{a6}
0 & = & \D_{b}\sigma^{ab} - (\c\omega)^{a}
- {\ts{2\over3}}\,\D^{a}\Theta + q^{a} \ , \\
\label{a8}
0 & = & (\c\sigma)_{ab} + \D_{\la a}\omega_{b\ra} - H_{ab}
\ , \\
\label{a9}
0 & = & \D_{b}E^{ab} - {\ts{1\over3}}\,\D^{a}\mu + H\,q^{a}
\ , \\
\label{a10}
0 & = & \D_{b}H^{ab} + {\ts{1\over2}}(\c q)^{a}
- \rho\,\omega^{a} \ ,
\end{eqnarray}
where $H = \dot{a}/a$ is the background Hubble expansion rate,
related to the background value $\rho$ of $\mu$ by
\begin{eqnarray}
\label{frw1}
H^{2} & = & {\ts{1\over3}}\,(\rho+\Lambda)
- {{\cal K}\over a^{2}} \ , \\
\label{frw}
\dot{H} & = & -{\ts{1\over2}}\,\rho + {{\cal K}\over a^{2}} \ .
\end{eqnarray}
The Friedmann equation (\ref{frw1}) may be rewritten in terms of
the dimensionless density parameters as
\begin{equation}
\label{frw1'}
1 = \Omega + \Omega_{\Lambda} + \Omega_{\cal K} \ .
\end{equation}
%



\end{document}